\documentclass[aps,prl,showpacs,preprint,letterpaper]{revtex4}
\usepackage{graphicx}
\usepackage{amsmath}
\begin{document}

\title{Dynamics and Thermodynamics of a Novel Phase of NaAlH$_4$}

\author{Brandon C. Wood\footnote{Present address: Lawrence Livermore National
        Laboratory, Livermore, CA 94550}}
\author{Nicola Marzari}
\affiliation{Department of Materials Science and Engineering, Massachusetts 
Institute of Technology, Cambridge, MA 02139}

\begin{abstract}
We characterize a novel orthorhombic phase ($\gamma$) of NaAlH$_4$, discovered 
using first-principles molecular dynamics, and discuss its relevance to the
dehydrogenation mechanism. This phase is close in energy to the known
low-temperature structure and becomes the stabler phase above 320~K, thanks to 
a larger vibrational entropy associated with AlH$_4$ rotational modes. The 
structural similarity of $\gamma$-NaAlH$_4$ to $\alpha$-Na$_3$AlH$_6$ suggests 
it acts as a key intermediate during hydrogen release. Findings are 
consistent with recent experiments recording an unknown phase during 
dehydrogenation.
\end{abstract}


\maketitle

Complex light-metal hydrides represent a materials solution to the 
hydrogen-storage problem~\cite{schlapbach01} because of their potential for 
acceptably high gravimetric capacities, as well as an ability to be engineered 
for favorable hydrogen release kinetics. Of these, sodium alanate 
(NaAlH$_{4}$) has attracted particular interest by combining relatively high 
theoretical hydrogen release with ready reversibility through the addition of 
transition metal dopants~\cite{bogdanovic97, schuth04}. As such, the material 
has been widely studied as a template for developing novel hydrogen storage 
solutions. Several recent first-principles studies~\cite{arroyo04, peles04, 
aguayo04, iniguez04, ozolins04, araujo05, lovvik05, ke05, vegge06, 
li06, frankcombe06, marashdeh06, peles06, du07, peles07, iniguez0705, 
gunaydin08} have addressed the structure and thermodynamics of the sodium 
alanate system. Nevertheless, important questions regarding the precise
mechanism of dehydrogenation and the nature of the key phase transition from 
NaAlH$_4$ to Na$_3$AlH$_6$ remain unanswered.

In this Letter, we present static, linear-response, and molecular dynamics (MD)
calculations based on density-functional theory (DFT) that support the 
existence of a hitherto unknown orthorhombic phase of NaAlH$_4$ ($\gamma$), 
found to be thermodynamically favored over the low-temperature $\alpha$ phase 
at relevant operating temperatures for dehydrogenation. This phase is 
entropically stabilized by activation of AlH$_4$ rotation and readily 
nucleates in MD simulations of (001) surface slabs. Our evidence is presented 
alongside recent experiments to highlight the likely significance of 
this phase in stabilizing and mediating the decomposition reaction.

Our calculations are based on DFT in the plane-wave pseudopotential
formalism~\cite{espresso} using the PBE exchange-correlation 
functional~\cite{pbe96}. Ultrasoft pseudopotentials~\cite{vanderbilt90} for Na 
(2\emph{s}$^{2}$2\emph{p}$^{6}$3\emph{s}$^{0.5}$), H, and Al 
(3\emph{s}$^2$3\emph{p}$^2$) were taken from the 
Quantum-Espresso distribution~\cite{espresso}. A second Troullier-Martins 
pseudopotential~\cite{tm91} for Na, used only for the MD runs, was 
generated in a 3\emph{s}$^{0.5}$3\emph{p}$^{0.05}$ valence configuration, with
nonlinear core corrections~\cite{louie82} added. 
For the MD simulations, cutoffs of 25 and 200~Ry were used for the 
wavefunctions and charge density, respectively, representing convergence of 
forces to within 0.02~eV/\AA. These were raised to 30 and
300~Ry for the total-energy and linear-response calculations.
The $\alpha \to \gamma$ transition was studied using Car-Parrinello MD 
simulations~\cite{cp85} of bulk supercells of NaAlH$_{4}$ with 96 atoms 
(2$\times$2$\times$1 $\alpha$ unit cells) and of slab supercells with 
192 atoms (8 layers; 2$\times$2$\times$2 $\alpha$ unit cells). Slabs were 
cleaved along the (001) plane---verified experimentally and theoretically as 
the stablest facet~\cite{frankcombe06}---with 14~\AA\ of vacuum inserted 
between images. Except where indicated, slab simulations were performed in the 
\emph{NVT} ensemble, whereas bulk simulations were performed in the \emph{NPT} 
ensemble at zero pressure using the Parrinello-Rahman extended Lagrangian
formalism~\cite{focher94}. The theoretical $\alpha$ lattice parameter was used 
for the initial cell state. Bulk NaAlH$_4$ was sampled at 50~K intervals from 
300--650~K; surface slabs at 25~K intervals from 225--425~K. Temperatures were 
maintained using Nos\'{e}-Hoover chains~\cite{martyna92}, with electronic 
fictitious mass $\mu = 500$~au and $\Delta t = 6$~au. Following 
5~ps of thermalization, each bulk simulation ran for 25~ps and each surface 
simulation for 15~ps.

\begin{figure}
        \centering
	\includegraphics[width=3.5in]{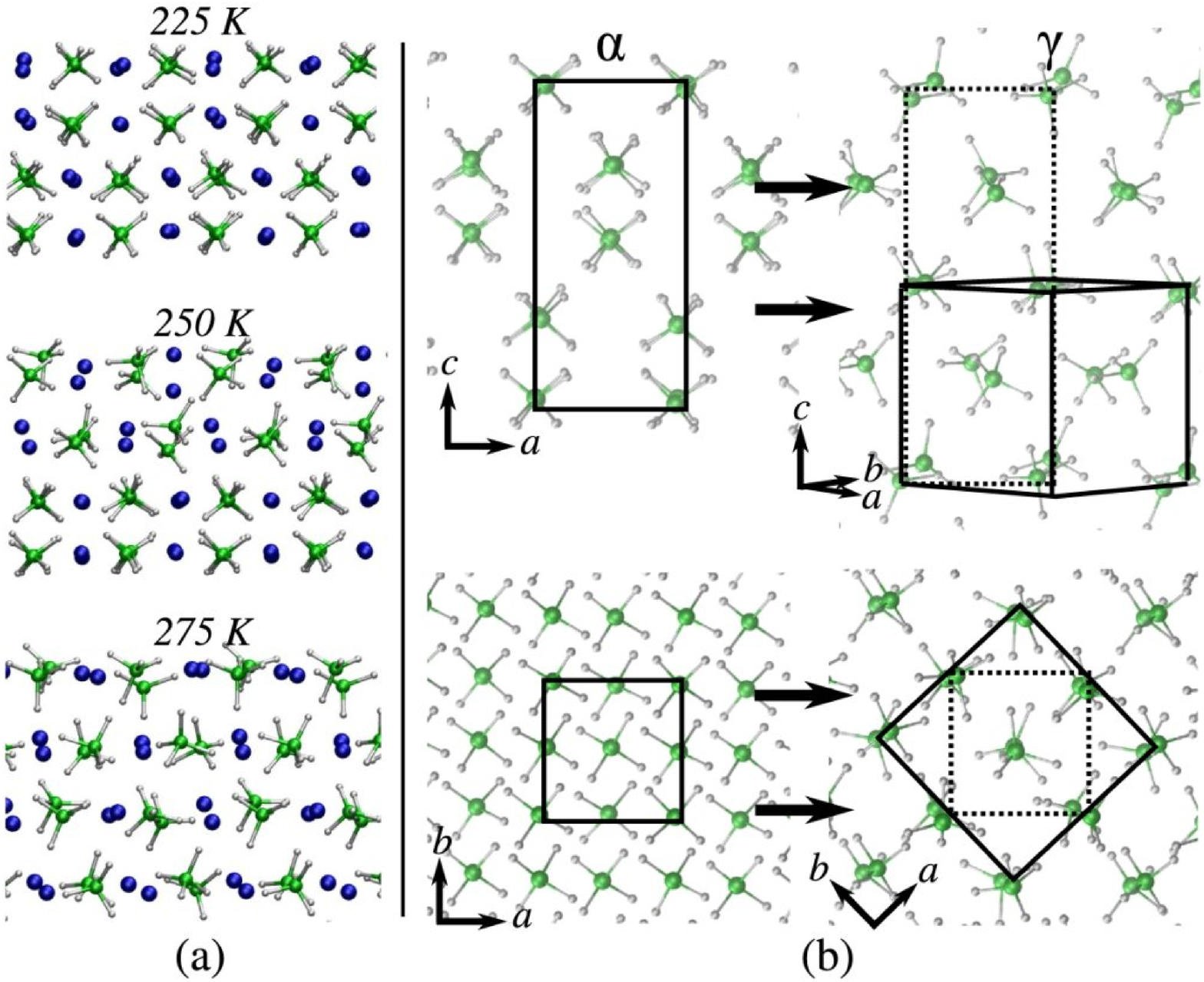}
	\caption
	{(Color online) (a) Equilibrated structure of an 8-layer (001) surface 
	slab of NaAlH$_{4}$ taken from MD runs at 225, 250, and 
	275~K (topmost four layers shown). Na/Al/H atoms are shown in 
	blue/green/white. Slabs are viewed along the $\mathbf{\hat{b}}$ axis 
	of the $\alpha$ phase, with $\mathbf{\hat{c}}$ oriented upwards. 
	Surface shear at 250~K signals the onset of $\gamma$, complete by
	275~K. (b) Schematic illustration of structural differences  
	before (left) and after (right) the transition (identical viewpoints, 
	Na atoms omitted). Solid lines outline the $\alpha$ and $\gamma$ 
	unit cells; dotted lines connect original boundary atoms of the 
	$\alpha$ cell.}
        \label{fig:alpha_to_gamma}
\end{figure}

The new phase, designated throughout this Letter as $\gamma$ to distinguish it 
from the high-pressure $\beta$ phase~\cite{kumar07}, manifests itself in our 
MD simulations of the (001) surface slab of NaAlH$_4$. Specifically, we find 
that the (001) slab exhibits a spontaneous transition from the $\alpha$ 
to the $\gamma$ phase for $250 < T < 350$~K, within 15~ps into the production 
run. This transition is clearly visible in Fig.~\ref{fig:alpha_to_gamma}a, 
which compares final configurations of the surface slab at select simulation 
temperatures. (Melting occurred above 350~K; tendency to melt is enhanced by 
our finite slab thickness). To rule out finite-size or supercell restrictions, 
we ran additional surface simulations at 300~K, one with a larger 270-atom 
supercell and one in the \emph{NPT} ensemble. Both evidenced the same 
transition to the $\gamma$ phase. An \emph{NPT} simulation of bulk 
$\gamma$-NaAlH$_4$ at 300~K was also performed, with the new phase remaining 
stable throughout the run.

Figure~\ref{fig:alpha_to_gamma}b illustrates the basic structural 
differences between the $\alpha$ and $\gamma$ phases, as observed
in the slab simulations. Three qualitative features of the $\alpha 
\to \gamma$ transition emerge in the MD runs: a rotational disordering of the 
AlH$_4$ groups; an expansion of the crystal along $\mathbf{\hat{c}}$; 
and a shear of successive crystal planes parallel to the surface, 
generating the lattice symmetry of the $\gamma$ phase. 

In contrast to the surface slab, no structural reorganization was observed in 
\emph{NPT} simulations of bulk NaAlH$_4$ (until melting at $T \geq 
550$~K). Given its facile expression in the (001) slab, we conclude that any 
observation of the transition to $\gamma$ in the bulk must be kinetically 
hindered, requiring much longer simulations for observation. This hindrance 
relates to the volume expansion of the lattice during the transition, as 
discussed further below.

\begin{figure}
        \centering
	\includegraphics[width=3.4in]{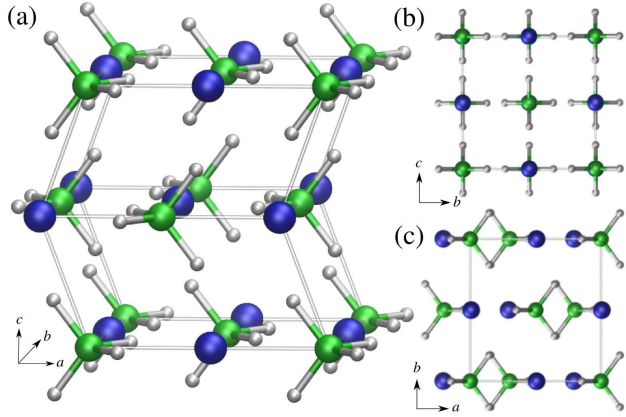}
        \caption
	{(Color online) Structure of orthorhombic $\gamma$-NaAlH$_{4}$, shown 
	(a) in the standard view; (b) along $\mathbf{\hat{a}}$; and (c) along 
	$\mathbf{\hat{c}}$. The color scheme follows 
	Fig.~\ref{fig:alpha_to_gamma}.}
        \label{fig:gamma}
\end{figure}

In Fig.~\ref{fig:gamma}, we present the structure of $\gamma$-NaAlH$_4$, taken
from representative dynamics timesteps and relaxed with respect to all 
internal parameters and lattice vectors. The unit 
cell is best described as a 12-atom base-centered orthorhombic (bco) structure 
with space group \emph{Pmmn} and $a=6.500$~\AA, $b=7.105$~\AA, and 
$c=7.072$~\AA\ (see Table~\ref{tab:gamma_pos} for details; note the 
near-degeneracy of $b$ and $c$). We compared the energy of this 
structure, calculated using 216 $\mathbf{k}$-points in the full Brillouin 
Zone, to that of the low-temperature $\alpha$ phase, calculated 
using a 324-$\mathbf{k}$ point mesh of similar density. At 0~K, the $\alpha$ 
phase is lower in energy by only 9~meV/atom, pointing to a close 
competition for stability between these two phases. The accuracy of the PBE
energetics was also checked against quantum-chemistry calculations on isolated
72-atom clusters of $\alpha$- and $\gamma$-NaAlH$_4$ using the B3LYP functional
and a 6-311+G basis; the difference between B3LYP and PBE in the relative 
stability of the two clusters (likely bracketing the exact result) was
2~meV/atom.

\begin{table}
	\caption{Internal atomic positions for $\gamma$-NaAlH$_4$ 
	(\emph{Pmmn}; $a=6.500$~\AA, $b=7.105$~\AA, $c=7.072$~\AA). 
	Units are fractional coordinates ($\alpha$, $\beta$, $\gamma$) of 
	the primitive bco lattice vectors.} 
	\label{tab:gamma_pos} 
\begin{tabular*}{0.45\textwidth} 
	{@{\extracolsep{\fill}}lccc} 
	\hline
	Atom type & ~$\alpha$~ & ~$\beta$~ & ~$\gamma$~ \\
	\hline
	Na & $\pm$0.141 & $\mp$0.141 & $\pm$0.250 \\
	Al & $\pm$0.349 & $\mp$0.349 & $\mp$0.250 \\
	H & $\pm$0.198 & $\mp$0.198 & $\mp$0.067 \\
	H & $\pm$0.198 & $\mp$0.198 & $\mp$0.433 \\
	H & $\pm$0.313 & $\pm$0.317 & $\mp$0.250 \\
	H & $\pm$0.317 & $\pm$0.313 & $\pm$0.250 \\
\end{tabular*}
\end{table}

A comparison of Figs.~\ref{fig:alpha_to_gamma} and \ref{fig:gamma} reveals 
differences between the $\gamma$ phase at finite temperature and 
its zero-temperature, fully relaxed equivalent. The zero-temperature structure 
features alignment of H (but not Al) in successive AlH$_4$ tetrahedra along 
$\mathbf{\hat{c}}$. At finite temperatures, $\gamma$-NaAlH$_4$ instead 
exhibits an average alignment of Al along $\mathbf{\hat{c}}$, a consequence of 
randomized AlH$_4$ orientations and increased configurational freedom.

\begin{figure}
    \centering
	\includegraphics[width=3.25in]{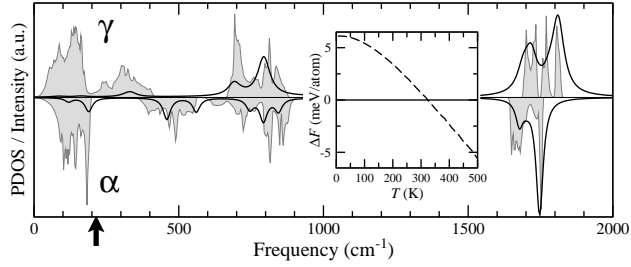}
    \caption
	{Phonon density of states (grey shaded region) and Raman spectrum 
	(solid black line, with thermal broadening) for $\gamma$- (top) and 
	$\alpha$-NaAlH$_4$ (bottom). Raman peaks below 1000~cm$^{-1}$ are 
	amplified tenfold for easier visibility. The arrow marks the 
	the $\alpha \to \gamma$ transition temperature. Inset shows the 
	temperature dependence of the free-energy difference $\Delta F = 
	F_{\gamma} - F_{\alpha}$.}
    \label{fig:phonon}
\end{figure}

To establish finite-temperature properties, we have calculated the
phonons~\cite{supplement} and phonon density of states (PDOS) for the $\alpha$ 
and $\gamma$ phases using density-functional perturbation theory 
\cite{baroni01}. Figure~\ref{fig:phonon} compares the PDOS for
$\alpha$-NaAlH$_4$ (in close agreement with Ref.~\onlinecite{peles06}) with
$\gamma$-NaAlH$_4$. Bands below 180~cm$^{-1}$ are Na/Al translational lattice 
modes; 180--580~cm$^{-1}$, AlH$_4$ rotational modes; 750--900~cm$^{-1}$, 
bending modes; and above 1600~cm$^{-1}$, stretching modes. The most
essential difference in the phonon spectra is that the AlH$_4$ 
rotational modes (350--580 cm$^{-1}$ in the $\alpha$ phase) are softened
substantially (200--420 cm$^{-1}$) in the $\gamma$ phase. This translates to an 
increase in vibrational entropy at temperatures for which these modes are 
active. We estimate the temperature-dependent free energy $F(T)$ of the 
phases by adding the Bose-Einstein vibrational entropy contribution onto the 
ground-state total energy $E_0$:
\begin{displaymath}
	F(T) = E_0 + \sum_{\mathbf{q},j} \left\{ \frac{1}{2} \hbar 
	\omega_{\mathbf{q},j} + k_\mathrm{B} T \ln \left[ 1 
   - \exp \left( -\frac{\hbar\omega_{\mathbf{q},j}}{k_\mathrm{B} T} 
	\right) \right] \right\}
\end{displaymath}
where sums run over band
indices $j$ and wavevectors $\mathbf{q}$ in the first Brillouin Zone.
Neglecting further contributions to $F(T)$, we find good agreement between
the transition temperature predicted by the free-energy analysis and that found
in the dynamics simulations (250~K) when using identical simulation 
parameters. Upon increasing the plane-wave cutoff and including the Na 
semicore states to more accurately account for the zero-point motion as 
reported in Ref.~\onlinecite{peles06}, we obtain the results in 
Fig.~\ref{fig:phonon}, indicating a thermodynamic preference for the 
$\gamma$ phase for $T \geq 320$~K (see inset). Notably, this establishes the 
$\gamma$ phase as the stablest variant within the experimental temperature 
range for dehydrogenation. 

\begin{figure}
        \centering
        \includegraphics[width=3.4in]{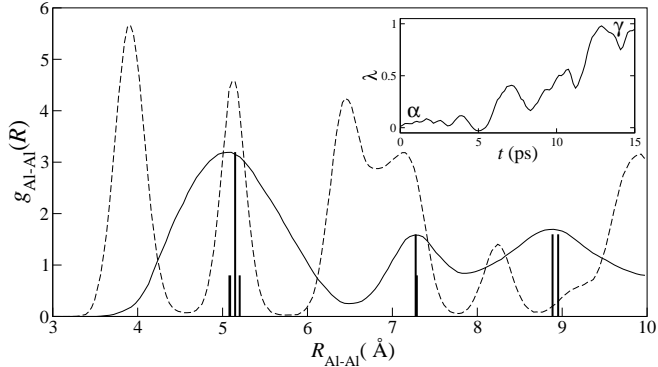}
        \caption[]
	{Comparison of the Al--Al pair correlation function 
	$g_\mathrm{Al-Al}$ for the (001) slab before (dashed line)
	and after (solid line) the $\alpha \to \gamma$ transition at 275~K. 
	Solid vertical lines give peak locations for a lattice-matched ideal 
	crystal of $\alpha$-Na$_3$AlH$_6$ (see text for definition), with
	relative peak heights shown for comparison. Inset shows the evolution 
	of the order parameter $\lambda$ during the transition.}
        \label{fig:gofr_orderparam}
\end{figure}

In a recent \emph{in-situ} Raman study~\cite{yukawa07}, Yukawa \emph{et al.} 
found evidence of an unknown intermediate phase of NaAlH$_4$ in first-stage 
dehydrogenation, emerging in the uncatalyzed sample just before melting at
458~K. The authors suggest that, like our proposed $\gamma$-NaAlH$_4$, this 
phase should be stable upon cooling to near room temperature and contain 
intact AlH$_4$ units with little constraint from the surrounding crystal 
lattice. Accordingly, we have calculated the Raman spectrum for 
$\gamma$-NaAlH$_4$, shown in Fig.~\ref{fig:phonon}~\cite{supplement}. A 
comparison between our results and the spectrum at 458~K in 
Ref.~\onlinecite{yukawa07} shows a remarkable agreement between the two. 
Foremost, we note the new signature Al--H stretching peak near 
1800~cm$^{-1}$ in both experiment and simulation. The simulations also record 
a second resonance at 1750~cm$^{-1}$; although not explicitly visible in
Ref.~\onlinecite{yukawa07}, asymmetry in their 1800~cm$^{-1}$ signature does 
suggest the presence of an additional lower-frequency peak. The difficulty in 
experimentally resolving these two peaks can be attributed to the thermally
averaged nature of the finite-temperature $\gamma$ structure with respect to 
its zero-temperature counterpart, as well as additional soft lattice modes 
activated during the transition (similar broadening is observable in the 
pair correlation data of Fig.~\ref{fig:gofr_orderparam}). In the experiment, 
the 400--500~cm$^{-1}$ peaks disappear, and distinct peaks at 770 and 
817~cm$^{-1}$ merge into a single broad peak. Both features are also visible 
in Fig.~\ref{fig:phonon}. Note that in reality, the 400--500~cm$^{-1}$ 
rotational modes in $\alpha$-NaAlH$_4$ do not disappear in $\gamma$ but rather 
shift to a broad peak at $\sim$300~cm$^{-1}$; however, the corresponding 
Raman-active peaks are too weak to demonstrate a resolvable signal. 
Collectively, these similarities present a strong case for the interpretation 
of the unknown phase in Ref.~\onlinecite{yukawa07} as $\gamma$-NaAlH$_4$.

A closer examination of the MD results elucidates why the $\alpha \to \gamma$ 
transition is not observed in our bulk \emph{NPT} simulations. In the surface 
slab simulations, $\gamma$-NaAlH$_4$ nucleated only upon activation of the 
rotational modes of surface AlH$_4$ units (center panel, 
Fig.~\ref{fig:alpha_to_gamma}a). This agrees with Fig.~\ref{fig:phonon}, since 
softening of AlH$_4$ rotational modes is the predominant contributor to the 
thermodynamic stability of the $\gamma$ phase. 
However, in bulk $\alpha$-NaAlH$_4$ the AlH$_4$ units cannot rotate: the 
barrier to rotational mobility is larger than for the relatively unconstrained 
surface groups, and rotation must be coupled with large volume fluctuations. 
This makes ready nucleation of the $\gamma$ phase in bulk NaAlH$_4$ difficult.
Correspondingly, rotation of interior AlH$_4$ units in the (001) surface slab 
is strongly coupled to the volume expansion perpendicular to the surface. 
The calculated surface energy of $\gamma$-NaAlH$_4$(001) is also much smaller 
than $\alpha$-NaAlH$_4$(001) (3.1 versus 8.2 meV/\AA$^2$), favoring 
the transition in the slab. The results show that AlH$_4$ rotation percolates 
once activated at the surface, as the slab thickness increases to 
accommodate. This lowers the barrier for $ab$-planar shear, finalizing the 
transition. We therefore assume that nucleation of $\gamma$-NaAlH$_4$ should 
occur more readily at an exposed (001) surface, at grain boundaries in a 
polycrystalline sample, or in low-stress regions that can accommodate the 
local volume fluctuations necessary for AlH$_4$ rotation. 

We investigate the transition timescale in the MD simulations by introducing 
an order parameter $\lambda$ as a reaction coordinate, defined as 
$\lambda = 1 - f_\mathrm{edge} / f_\mathrm{corner}$. Here $f_\mathrm{edge}$ 
and $f_\mathrm{corner}$ are the fractional occupancies of the edge-center and 
corner sublattices in the projection of the $\alpha$ unit-cell geometry onto 
the surface plane (see Fig.~\ref{fig:alpha_to_gamma}b). The rapid evolution 
from $\lambda=0$ (i.e., pure $\alpha$) to $\lambda=1$ (i.e., pure $\gamma$) 
evidences the strong thermodynamic driving force for the transition (see 
inset, Fig.~\ref{fig:gofr_orderparam}).

Geometric similarities in the aluminum substructures of $\gamma$-NaAlH$_4$ and
$\alpha$-Na$_3$AlH$_6$ suggest a phenomenological connection between the 
$\alpha \to \gamma$ transition and the NaAlH$_4$ $\to$ 
Na$_3$AlH$_6$ dehydrogenation reaction. These are quantified in 
Fig.~\ref{fig:gofr_orderparam}, which compares the Al--Al pair correlation 
function for the (001) slab during the $\alpha \to \gamma$ transition 
with one for Na$_3$AlH$_6$ that is constrained to be lattice matched to the 
$\gamma$ cell. Specifically, $\alpha$-Na$_3$AlH$_6$ peaks were obtained by 
isotropically shrinking the cell volume (-10\%) to match the mean lattice 
parameter in the $ab$ plane to the $\gamma$-NaAlH$_4$ cell boundaries.
The striking agreement between the locations and relative heights of the two 
sets of peaks strongly hints at the role of $\gamma$-NaAlH$_4$ as a key 
intermediate during dehydrogenation. If so, the role of exposed surfaces in 
nucleating $\gamma$ may help to account for the enhancement of dehydrogenation 
kinetics due to ball milling without a catalyst~\cite{anton03}, as 
well as improved kinetics of NaAlH$_4$ nanoparticles~\cite{zheng08}. We
did not examine the impact of transition metal additives; however, 
the catalyst is likely to further aid the conversion of $\gamma$-NaAlH$_4$ to 
$\alpha$-Na$_3$AlH$_6$. A detailed investigation of this possibility is 
currently underway.

In conclusion, we have discovered and characterized a new phase of NaAlH$_4$,
labeled here as $\gamma$. Our description is fully consistent with an 
unknown intermediate found in a recent \emph{in-situ} Raman study. Activation
of the AlH$_4$ rotational modes drives the transition to $\gamma$, which 
occurs readily in (001) surface slab simulations for $T \geq 320$~K. 
The thermodynamic stability of $\gamma$-NaAlH$_4$ beyond this 
temperature has been further confirmed by vibrational entropy calculations 
using density-functional perturbation theory. Remarkable structural 
similarities between $\gamma$-NaAlH$_4$ and $\alpha$-Na$_3$AlH$_6$ suggest 
that the $\gamma$ phase should play a key role in mediating dehydrogenation. 
Notably, no Al--H bonds were broken in our simulations; this agrees with 
Ref.~\onlinecite{majzoub05} and underscores the need for an additional 
defect-driven mechanism to fully account for dehydrogenation. Accordingly, 
further investigation of low-energy defects in $\gamma$-NaAlH$_4$ that could 
facilitate Al--H cleavage is recommended.

Funding was provided by DOE via CSGF and Hydrogen Program 
DE-FG02-05ER46253. Calculations were performed using 
Quantum-Espresso~\cite{espresso} on facilities provided through NSF Grant 
DMR-0414849. The authors thank Gerbrand Ceder for helpful discussions, and 
Elise Li for the quantum-chemistry results.

\end{document}